# Nematic liquid crystal boojums with handles on colloidal handlebodies

Qingkun Liu[1], Bohdan Senyuk[1,2], Mykola Tasinkevych[3,4], and Ivan I. Smalyukh[1,2,5,6,*]

[1]*Department of Physics, University of Colorado, Boulder, CO 80309*

[2]*Liquid Crystal Materials Research Center, University of Colorado, Boulder, CO 80309*

[3]*Max-Planck-Institut für Intelligente Systeme, D-70569 Stuttgart, Germany*

[4]*Institut für Theoretische Physik IV, Universität Stuttgart, D-70569 Stuttgart, Germany*

[5]*Department of Electrical, Computer, and Energy Engineering and Materials Science and Engineering Program, University of Colorado, Boulder, CO 80309;*

[6]*Renewable and Sustainable Energy Institute, National Renewable Energy Laboratory and University of Colorado, Boulder, CO 80309*

*E-mail: ivan.smalyukh@colorado.edu

## Abstract

Topological defects that form on surfaces of ordered media, dubbed boojums, are ubiquitous in superfluids, liquid crystals (LCs), Langmuir monolayers, and Bose–Einstein condensates. They determine supercurrents in superfluids, impinge on electrooptical switching in polymer-dispersed LCs, and mediate chemical response at nematic-isotropic fluid interfaces, but the role of surface topology in the appearance, stability, and core structure of these defects remains poorly understood. Here, we demonstrate robust generation of boojums by controlling surface topology of colloidal particles that impose tangential boundary conditions for the alignment of LC molecules. To do this, we design handlebody-shaped polymer particles with different genus $g$. When introduced into a nematic LC, these particles distort the nematic molecular alignment field while obeying topological constraints and induce at least $2g - 2$ boojums that allow for topological charge conservation. We characterize 3D textures of boojums using polarized nonlinear optical imaging of molecular alignment and explain our findings by invoking symmetry considerations and numerical modeling of experiment-matching director fields, order parameter variations, and nontrivial handle-shaped core structure of defects. Finally, we discuss how this interplay between the topologies of colloidal surfaces and boojums may lead to controlled self-assembly of colloidal particles in nematic and paranematic hosts, which, in turn, may enable reconfigurable topological composites.

# Introduction

Being inspired by Lewis Carroll's poem "The Hunting of the Snark," Mermin (1, 2) introduced a term "boojum" to name elusive at the time surface defects in superfluids. This term and the concept of boojums quickly penetrated different fields of physics and materials science, ranging from liquid crystals (LCs) to Langmuir monolayers, and to Bose–Einstein condensates (3–6). However, unlike in the case of their bulk topological counterparts (7–10), called "hedgehog" point defects, the appearance of boojums is rarely controlled at will. In LCs, boojums can spontaneously appear at their interfaces with isotropic media, such as water. They are commonly associated with the geometries of thin LC films on surfaces of isotropic fluids (11), LC droplets (3, 4, 12), and colloidal inclusions (13, 14). For example, spherical surfaces with tangential boundary conditions for rod-like LC molecules, and director **n** describing their local average orientation, are known to induce two boojums per LC droplet or per colloidal inclusion in the LC host (3, 4, 13, 14). However, despite the recent progress in generating and controlling bulk LC defects using director realignment with focused laser beams, chirality, and surface topology of colloidal particles with perpendicular surface anchoring introduced into LCs (7–10), similar control of boojums has not been demonstrated.

In this work, we fabricate polymer microparticles with the topology of a handlebody of genus *g* ranging from one to five that are capable of imposing tangentially degenerate surface boundary conditions for the director **n**. Using a combination of holographic laser tweezers (HOT), bright-field microscopy, polarizing optical microscopy (POM), and three-photon excitation fluorescence polarizing microscopy (3PEF-PM) (15, 16), we control and deduce multistable topology-constrained director patterns that form around these handlebody-shaped particles. We show that 2D defects in orientation of the director at the LC–colloid interface have a net strength adding to the value of the Euler characteristic of the handlebody particle $\chi = 2 - 2g$, as expected (17), although one typically also encounters defects of opposite signs at different locations on handlebody surfaces that self-compensate each other and often annihilate with time, providing multiple different ways of satisfying topological constraints at multiple stable or metastable orientations of particles. Topological charges of 3D textures of boojums, on the other hand, satisfy the topological charge conservation due to all defects induced in the LC. Numerical modeling based on free energy minimization yields results consistent with experimental findings and provides insights into different handle-shaped core structures of the corresponding boojums. We discuss these findings from the standpoint of self-assembly–based approaches for fabrication of a unique class of topological composites.

## Results and Discussion

The lithographically fabricated handlebody colloids (details are provided in *Materials and Methods* and Figs. S1 and S2) had a surface Euler characteristic $\chi = 0, -2, -4, -6, -8$ (Table 1); lateral diameters ranging from 5 to 10 µm; and rounded-square cross-sections of about 1 µm. A dispersion of these particles in a nematic LC, pentyl cyanobiphenyl (5CB), was infiltrated into thin cells, which, in the absence of inclusions, had a uniform in-plane director $n_0$ in their interior. The handlebody colloids induced director distortions around them that are described by the director field $n(r)$ approaching $n_0$ at large distances. To minimize free energy due to elastic distortions (18), the particles tend to align with their ring planes parallel to $n_0$ (Fig. 1), although metastable configurations with ring planes perpendicular to $n_0$ are occasionally observed (Fig. 2) and can be reproducibly obtained using local melting of the LC by laser tweezers, followed by quenching it back to the nematic state. Although mass density of the ultraviolet-sensitive polymer SU-8 (~1,190 kg/m$^3$) comprising particles is slightly higher than that of the 5CB nematic host (~1,020 kg/m$^3$), repulsive elasticity-mediated interactions between particles and confining surfaces of cells with strong planar surface anchoring of $n(r)$ cause levitation of the colloidal tori in both orientations in the vicinity of cell midplane (Figs. 1 F and G and 2E) while undergoing both translational and rotational thermal fluctuations.

A single colloidal torus with a ring plane parallel to $n_0$ is found to induce four boojums (Fig. 1). A less frequently observed metastable configuration of a torus particle aligned perpendicular to $n_0$ contains no boojums (Fig. 2) but, instead, a nonsingular axially symmetrical $n(r)$ (with bend, splay, and twist distortions) that satisfies tangential boundary conditions on the particle surface while approaching $n_0$ at large distances from its surface (Fig. 2 G–I). The pattern of the 2D "surface" nematic director $n_s(r)$ at the surface of a torus with tangentially degenerate boundary conditions contains no defects in the latter case but four 2D defects (point disclinations) in the former case, two of strength $s = 1$ and two of strength $s = -1$, where $s$ is defined as the number of times $n_s(r)$ rotates by $2\pi$ as one circumnavigates the defect core once (Fig. 1H). Numerical calculation of Landau–de Gennes free energy for rings parallel and perpendicular to $n_0$ reveals that the latter is a metastable state with a local minimum of the free energy (Fig. 2F). Rotation of the particle away from the local-minimum perpendicular orientation by 5–10° results in nucleation of defects and subsequent reorientation to the ground-state configuration with the ring plane along $n_0$; it is this free energy barrier set by the energetic cost of boojum nucleation that stabilizes the metastable configuration against realignment to the ground state. This behavior is consistent with the numerical calculation of the elastic free energy at different fixed orientations of the colloidal torus with respect to $n_0$ (Fig. 2F). Importantly, for particles oriented perpendicular to $n_0$, $n(r)$ twists and then untwists in an axially symmetrical fashion to meet tangential boundary conditions at the particle surface and ensure that $n(r)$ is parallel to $n_0$ both in the center of the torus and far from the particle in its exterior (Fig. 2 G and H). Furthermore, director configurations having opposite senses of twist in both the interior and exterior of the ring (one of which is shown in Fig. 2 G and H) are observed with equal probability.

Handlebody colloids with $g > 1$ aligned with ring planes parallel to $n_0$ induce several stable and metastable configurations of $n(r)$ with different numbers and locations of boojums. This type of behavior of handlebodies with $g = 2$ is illustrated in Fig. 3 and Fig. S3. The number of induced boojums in this case of $g = 2$ is always even and typically varies from 6 to 10 (Table 1). As well as two topologically required $s = -1$ defects in $n_s(r)$, one observes additional self-compensating defects of opposite strengths. Only a small number out of many metastable configurations is presented in Fig. 3 and Fig. S3; the other configurations (including metastable configurations for particles of $g > 1$ oriented perpendicular to $n_0$) will be discussed in detail elsewhere. The total strength of disclinations in $n_s(r)$ is always $\Sigma_i s_i = -2$ for handlebodies of $g = 2$ (Table 1), as required to satisfy topological constraints (17). Various types of alignment and defect configurations are possible (Fig. 3), including ones where the axis-connecting centers of two rings align parallel, perpendicular, or at several oblique angles to $n_0$. One often observes diffusion of boojums along the particle surface and annihilation of defects with opposite signs, such that the number of surface defects typically decreases with time but the total strength remains constant, $\Sigma_i s_i = \chi = -2$. Both thermally driven diffusion and pinning of defects at surface imperfection occasionally give rise to locations of defects somewhat deviating from the ones in simulated textures (e.g., compare Fig. 3 B, E, and J, showing a stronger asymmetry of textures in simulations than in the experiments). Similar conclusions can be extended to colloids of genus 3, 4, and 5 (Figs. 4 and 5 and Table 1). For all colloids, the Euler characteristic measures the total strength of the point disclinations piercing the LC–colloid interface, $\Sigma_i\, s_i = \chi$, in agreement with the Poincaré–Hopf index theorem (17). Although some boojums of opposite signs annihilate with time and director structures around particles transform between different metastable and stable states, this relation always prevails. For example, two of the three boojums shown in the upper part of Fig. 4O (with surfaces of reduced scalar-order parameter depicted using red and green colors) can annihilate with time, in agreement with the experiments (Fig. 4 A and B). However, defect annihilation typically does not proceed until a minimum number of topologically required boojums is left, because some of these defects help to reduce elastic free energy (Figs. 1–5 and Table 1). Furthermore, colloids in all stable and metastable states undergo anisotropic diffusion (18), exhibiting both rotational and translational thermal motion (Fig. 5 D and E) dependent on the particular director and defect configurations induced by the colloids.

To characterize the reconstructed $\mathbf{n}(\mathbf{r})$ around a boojum, we first surround it by a hemisphere $\sigma$ on the LC side and calculate a characteristic $m_b = (1/4\pi)\int_\sigma d\theta d\phi\, \mathbf{n} \cdot \left[\frac{\partial \mathbf{n}}{\partial \theta} \times \frac{\partial \mathbf{n}}{\partial \phi}\right]$, where $\theta$ and $\phi$ are arbitrary coordinates on the hemisphere surface surrounding the boojum and, despite the medium's nonpolar nature, we treat $\mathbf{n}(\mathbf{r})$ as a vector field (Fig. 6 A–D). For simplicity, we assume that the surface of the particle in the vicinity of the boojum is locally flat, which is justified by the fact that the size of colloids is much larger than that of boojum cores, such that the radius of the hemisphere $\sigma$ can be selected to be much smaller than the radius of local curvature associated with the particle's surface. Within this approximation, boojums induced by colloids with tangential boundary conditions always have $m_b = \pm 1/2$ (Fig. 6 E–H) and add to zero, $\Sigma_i\, m_{bi} = 0$ (Fig. 6 and Table 1). We then calculate an integer "bulk charge" $N_b$ (4, 19, 20) while ensuring continuity of the "vectorized" $\mathbf{n}(\mathbf{r})$, which is $N_b = m_b + s/2$ for the tangential surface boundary conditions used in our experiments. We find that $\Sigma_i\, N_{bi} = \pm\chi/2$ for all studied colloids of $g = 1$–5 (Table 1), consistent with findings of our earlier studies for bulk point defects induced by colloidal handlebodies with perpendicular surface anchoring (8). We note that the bulk charge $N_b$ has a meaning not restricted to boojums but also describes bulk point defects and disclination loops. An experimental result $\Sigma_i\, N_{bi} = \pm\chi/2$ does not simply derive from relations $N_b = m_b + s/2$, $\Sigma_i\, m_{bi} = 0$, and $\Sigma_i\, s_i = \chi$ but describes conservation of the net bulk topological charge due to colloids and colloid-induced defects, similar to that known for bulk defects (8). The signs of all defects can be changed to the opposite signs by reversing the direction of the vectorized **n(r)** (8), as indicated by presence of ± in the relations above, but their relative charges and charge conservation always prevail.

The four different types of boojums experimentally observed in our system are schematically shown in Fig. 6 E–H. These boojums are of hyperbolic type and can be thought of as bulk hyperbolic defects cut into halves by different mirror symmetry planes coinciding with the colloidal surfaces. The other halves, in fact, can be thought of as virtual parts of these point defects located within the volume of the particles themselves, as depicted for the $g = 1$ torus in Fig. 6C. This allows us to understand the experimentally observed relation $\Sigma_i\, m_{bi} = 0$ simply as conservation of point defect charges, whereas the fact that the bulk defect charges $N_b$ of all boojums add to the value of the Euler characteristic, $\Sigma_i\, N_{bi} = \pm\chi/2$, is again consistent with the Gauss–Bonnet theorem and our previous studies (8, 12). Just like in the case of bulk point defects induced by colloids with normal boundary conditions (8), understanding the conservation of topological charges in the 3D textures (including both the real and virtual parts shown in Fig. 6C) is impossible without introducing the base point (8) and decorating **n(r)** with a vector field until the relative charges of all defects are determined. In addition to satisfying topological constraints on overall charge and 3D **n(r)** (8), the boojums satisfy the constraints on the overall strength of disclinations in **n$_s$(r)** at the particle–LC interface (Table 1).

Numerical modeling reveals different core structures of boojums associated with opposite strengths $s = \pm 1$ of point disclinations in **n$_s$(r)** (Figs. 3 K and L, 4 O–Q, and 6 I and J). The $s = \pm 1$ defects in **n$_s$(r)** locally split into pairs of half-integer disclinations of equivalent total strength of singularity. By plotting equilibrium isosurfaces of the reduced scalar-order parameter $Q = 0.25$ and **n(r)** in the LC bulk in the vicinity of boojums, we find that, rather interestingly, these split-core boojums (21) are composed of handle-shaped disclination semiloops terminating at the ±1/2 defects in **n$_s$(r)** (Figs. 3 K and L, 4 O–Q, and 6 I and J). Similar handle-shaped cores are expected to exist in all hyperbolic boojums at surfaces with strong surface anchoring (21) and are not related to the fact that the used colloids have handles. Furthermore, this splitting of cores of boojums into semicircular disclination loops is reminiscent of bulk point defects having a core structure in the form of handle-shaped disclination loops (21–28). Because of decreased $Q$ in the cores of disclination semiloops, which can be treated as being biaxial or isotropic in nature (3, 18, 29), one can define additional topologically nontrivial surfaces separating defect cores and the rest of the LC. The genus due to all surfaces, $g_{all}$, is thus the genus $g$ of the particle itself plus the number of disclination handles due to cores of all boojums induced by the colloid (Table 1), which is at least $2g - 2$, yielding high-genus surfaces with $g_{all} \geq 3g - 2$. By reducing the strength of surface anchoring or the overall size of colloids, one can also obtain boojums with nonsplit cores (21), thus controlling the surface defect topology and $g_{all}$.

To conclude, we have introduced a means for robust control of surface point defects at nematic-colloidal interfaces using surface topology that allows for generation of structures with quantized topology-constrained numbers of boojums dependent on particle's genus. Because the LC alignment at nematic-isotropic interfaces is sensitive to the presence of various biological molecules and microbes (30–32), our findings may enable development of bio-detectors and chemical sensors based on particles or droplets with topologically distinct surfaces and detection modes involving only local transformation of structures with different quantized numbers of boojums instead of the typical use of the global realignment of $\mathbf{n}_s(\mathbf{r})$ (30). The sensitivity may be dramatically enhanced in this case because a smaller number of biological molecules would be needed to induce transitions between various metastable and stable structures associated with comparable elastic free energies and satisfying topological constraints. It will also be of great interest to explore elasticity-mediated interactions and self-assembly (33–37) of colloidal particles with handles because, to date, such interactions have been studied only for particles with $g = 0$. In addition to elastic interactions, the mesomorphic fluid host brings another intriguing possibility of defect-mediated "colloidal bonding" (similar to that of atoms) associated with defects in $\mathbf{n}_s(\mathbf{r})$. In the case of paranematic ordering in an isotropic phase of the LC near isotropic-nematic transition temperature (38), when the scalar-order parameter is finite at particle surfaces but approaches zero in the bulk of colloidal dispersion, the surface defects in $\mathbf{n}_s(\mathbf{r})$ are all expected to be half-integer disclinations emanating from the particle surfaces into the bulk. Although various stable and metastable states can be expected, just like in the nematic LC bulk, topology will dictate the presence of at least $2|\chi|$ half-integer defects of total strength equal to $\chi$. Bonding of these defects emanating from the surfaces of different particles may thus enable unique types of topological defect-mediated self-assembly, similar to that previously proposed for spheres (38, 39). Colloidal particles with a different genus will allow for having different topologically constrained numbers of defect bonds, thus paving the road to unique modes of topology-assisted self-assembly in soft matter. Future studies will also explore dynamics of nucleation and annihilation of defects as particles are continuously rotated by external magnetic and optical torques, the influence of elastic constant anisotropy on the stability of different stable and metastable configurations, and hysteresis in switching between different states, as well as how spatial localization of boojums is related to intrinsic curvature of colloidal surfaces. The experimental arena we have developed, combined with the ability to achieve optical control of surface anchoring on colloidal surfaces (40) and the use of chiral LCs (7), may become a test ground for applying and probing a number of newly developed theories describing the topology of nematic defects, such as the Čopar–Žumer theory (41), and may be extended to topologically nontrivial LC confinement geometries (12).

## Materials and Methods

**Particle Fabrication.** Microfabrication of colloidal polymer handlebodies involved the following procedure (Fig. S1). First, a 1-μm-thick $SiO_2$ layer was deposited on a silicon wafer using plasma-enhanced chemical vapor deposition. Second, a 1.5-μm-thick layer of SU-8 photoresist (Microchem) was spin-coated onto the $SiO_2$ layer, and the pattern of rings was defined by standard photolithographic technique (8). The $SiO_2$ layer was then wet-etched in a buffered solution ($HF/NH_4F/H_2O$ = 3:6:10 by weight), and the handlebody particles made of SU-8 polymer were released into the solution. After washing in deionized water and redispersing these particles in 5CB, the nematic dispersion was infiltrated into cells composed of indium-tin-oxide–coated glass plates separated by glass spacers defining the cell gap. Cell substrates were coated with polyimide PI2555 (HD Microsystems) for in-plane alignment of $n_0$ defined by rubbing.

Optical manipulation and 3D imaging of samples were performed with an integrated setup of HOT and 3PEF-PM (15, 16) built around an IX 81 inverted microscope (Olympus) and using a 100× oil-immersion objective with an N.A. of 1.4. HOT used a phase-only spatial light modulator (Boulder Nonlinear Systems) and an Ytterbium-doped fiber laser (IPG Photonics) operating at 1,064 nm. 3PEF-PM used a tunable (680–1,080 nm) Ti-Sapphire oscillator (Coherent) emitting 140-fs pulses at a repetition rate of 80 MHz and an H5784-20 photomultiplier tube detector (Hamamatsu).

Bright-field and POM images (obtained both with and without a phase retardation plate) provide information about the location of boojums with respect to colloids, as well as a 2D "view" of a complex 3D $n(r)$ around the handlebodies. POM images (Figs. 1B and 2B) obtained with a full-wave retardation (530 nm) plate having the "slow axis" $Z'$ inserted after the planar LC cell exhibit textures with colors directly related to the orientation of **n(r)** (which is also the optical axis) with respect to $Z'$. When the phase retardation due to the sample is small, as in the case of LC cells with **n(r)**-distortions induced by handlebodies in planar cells, the blue-colored sample regions in Fig. 1B correspond to having **n(r)** at orientations roughly parallel to $Z'$, whereas the yellow color corresponds to **n(r)**$\perp Z'$.

The intrinsic 3D optical resolution of the nonlinear 3PEF-PMimaging due to the highly localized multiphoton excitation and strong ($\propto \cos^6 \alpha$) dependence of the 3PEF-PM fluorescence on the angle $\alpha$ between **n(r)** and the linear polarization of excitation light (15, 16) allows for the reconstruction of complex 3D director fields around handlebodies. The maximum-intensity areas in the fluorescence textures correspond to the linear polarization of the laser excitation light being parallel to anisotropic LC molecules and **n(r)**. The dark areas with minimum fluorescence are observed when polarization of excitation light is roughly perpendicular to the LC director. Color-coded 3PEF-PM images for different mutually orthogonal polarizations of laser excitation were overlaid on top of each other to obtain the images shown in Figs. 1D and 3G.

**Numerical Modeling.** We describe nematic LCs by a traceless symmetrical order-parameter tensor $Q_{ij}$, $i, j = 1, \ldots, 3$, and the corresponding Landau–de Gennes free energy functional (18):

$$F_{\text{LdG}} = \int_\Omega (f_b + f_{el}) d^3 x + \int_{\partial \Omega} f_s \, ds, \tag{1}$$

where $f_b$ and $f_{el}$ are the bulk and elastic free energy densities, respectively:

$$f_b = a \operatorname{Tr} \mathbf{Q}^2 - b \operatorname{Tr} \mathbf{Q}^3 + c \left( \operatorname{Tr} \mathbf{Q}^2 \right)^2, \tag{2}$$

$$f_{el} = \frac{L_1}{2} \partial_k Q_{ij} \partial_k Q_{ij} + \frac{L_2}{2} \partial_j Q_{ij} \partial_k Q_{ik}. \tag{3}$$

Here, $a$ is assumed to depend linearly on the temperature $T$, and $b$ and $c$ are considered temperature-independent constants. $L_1$ and $L_2$ in Eq. 3 are phenomenological parameters that can be related to the Frank–Oseen elastic constants (18, 26). The first integral in Eq. 1 is over the 3D domain, $\Omega$. The second integral in Eq. 1 accounts for finite surface anchoring conditions and is over the surfaces $\partial\Omega$ of the colloidal particles. We model planar degenerate anchoring at the colloidal surfaces using the surface potential (21–23):

$$f_s = W_1\left(\tilde{Q}_{ij} - \tilde{Q}_{ij}^\perp\right)^2 + W_2\left(\tilde{Q}_{ij}^2 - \left(\frac{3Q_b}{2}\right)^2\right)^2, \tag{4}$$

where $\tilde{Q}_{ij} = Q_{ij} + Q_b \frac{\delta_{ij}}{2}$, $\tilde{Q}_{ij}^\perp = (\delta_{ik} - v_i v_k)\tilde{Q}_{ij}(\delta_{lj} - v_l v_j)$, and $v$ is the normal to the colloidal surface. $W_1 > 0$ is the anchoring strength favoring the tangential orientation of $\mathbf{n}_s(\mathbf{r})$, and $W_2 > 0$ ensures the existence of a minimum for the surface scalar-order parameter at bulk value $Q_b$ (SI Text). For simplicity, we assume $W_1 = W_2 = W$. We minimize the Landau–de Gennes free energy (Eq. 1) numerically by using adaptive finite elements methods. More details can be found in SI Text and in the studies by Tasinkevych et al. (21) and Schopohl and Sluckin (42).

**Acknowledgments**

We thank M. Bowick, R. Budney, N. A. Clark, R. D. Kamien, R. B. Kusner, O. D. Lavrentovich, T. C. Lubensky, M. Ravnik, and S. Žumer for valuable discussions. This work was supported by National Science Foundation Grants DMR-0847782 (to Q.L. and I.I.S.), DMR-0820579 (to B.S. and I.I.S.), and DMR-0844115 (to I.I.S.), and partially by 7th Framework Programme International Research Staff Exchange Scheme Marie-Curie Grant PIRSES-GA-2010-269181 (to M.T.).

**Figures**

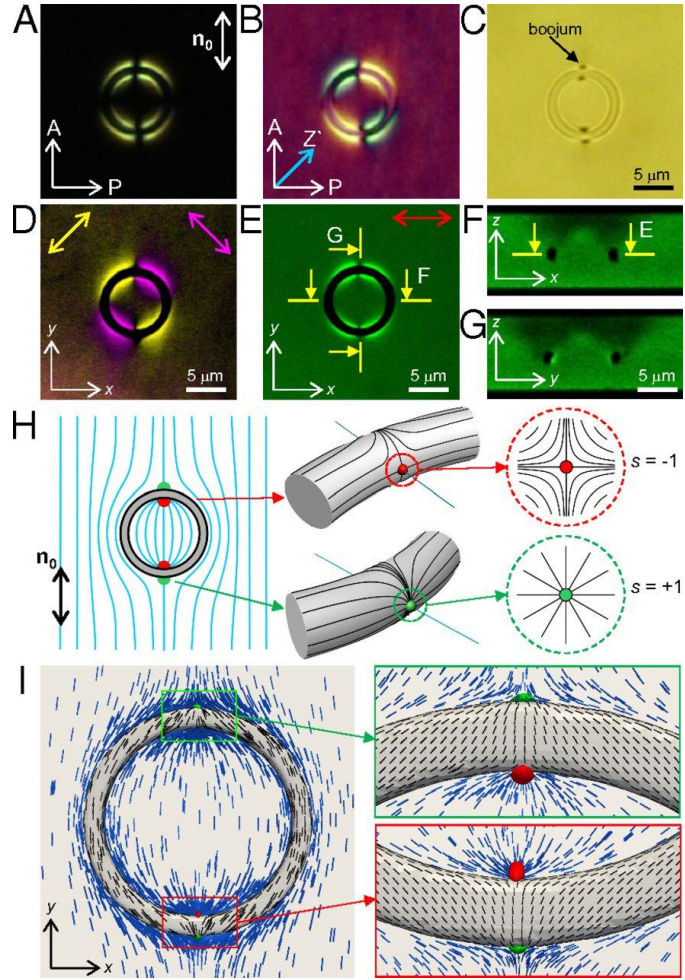

**Fig. 1.** Colloidal torus with tangential anchoring aligned parallel to $\mathbf{n}_0$. POM textures taken without (A) and with (B) a full-wavelength (530 nm) retardation plate with a slow axis $Z'$. (C) Bright-field microscopy texture of a torus with four induced boojums visible as dark spots. 3PEF-PM images of the same torus obtained for in-plane (D and E) and vertical (F and G) cross-sections, where the image (D) is a superimposition of 3PEF-PM textures taken for two orthogonal polarizations of laser excitation marked by yellow and magenta double arrows. The linear 3PEF-PM polarization direction in E–G is marked by a red double arrow in E. The mutual relative locations of cross-sections in E–G are marked on these images. (H) Schematic diagram of $\mathbf{n}(\mathbf{r})$ distorted by the handlebody immersed in a sample with a uniform far-field director. (*Insets*) Defects in $\mathbf{n}_s(\mathbf{r})$ on its surface. Green and red semispheres and circles represent, respectively, radial ($s = +1$) and hyperbolic ($s = -1$) surface disclinations in $\mathbf{n}_s(\mathbf{r})$. (I) Numerically calculated $\mathbf{n}(\mathbf{r})$ in the bulk (blue rods) and $\mathbf{n}_s(\mathbf{r})$ on the surface (black rods) of a colloidal torus. (Insets) Details of $\mathbf{n}(\mathbf{r})$ and $\mathbf{n}_s(\mathbf{r})$ near boojums and the isosurfaces of constant reduced scalar order parameter $Q = 0.35$ are shown by green and red colors, respectively.

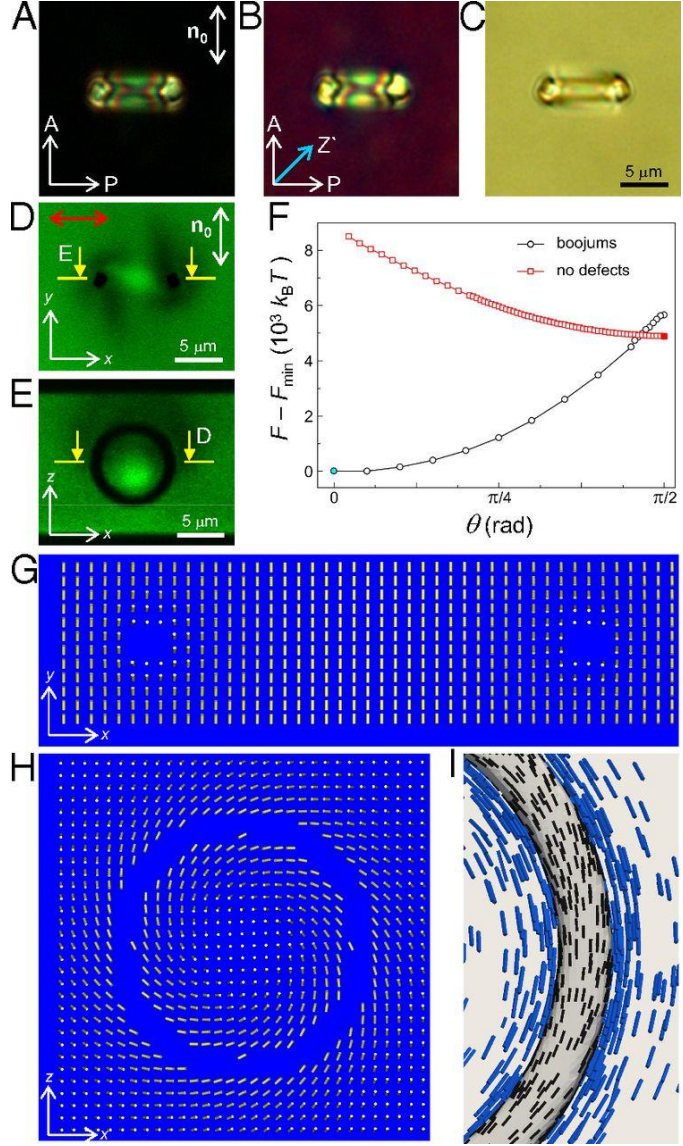

**Fig. 2.** Colloidal torus aligned with the ring plane perpendicular to $\mathbf{n}_0$ in a planar LC cell. POM textures taken without (A) and with (B) a retardation plate. (C) Bright-field microscopy texture reveals the absence of boojums. 3PEF-PM textures of in-plane (D) and vertical (E) cross-sections of the same torus; the yellow arrows mark the relative locations of cross-sections on the images. (F) Landau–de Gennes free energy (in units of $k_B T$, where $k_B$ is the Boltzmann constant and $T$ is the absolute temperature) vs. the angle between the ring plane of the torus and $\mathbf{n}_0$ for structures without (red open squares) and with (black open circles) boojums; the blue-filled circle corresponds to the ground-state orientation shown in Fig. 1 (ring parallel to $\mathbf{n}_0$), and the red-filled square corresponds to the metastable orientation shown in A (ring perpendicular to $\mathbf{n}_0$). Note that the existence of a boojum-free metastable $\mathbf{n}(\mathbf{r})$ at all angles $\theta > 5°$ and boojum-containing $\mathbf{n}(\mathbf{r})$ at $\theta < 89°$ is expected to cause hysteresis in structural behavior as particles are continuously rotated from $\theta = 0°$ to $\theta = 90°$ and back. (G and H) Numerically calculated $\mathbf{n}(\mathbf{r})$ (shown by cylinders) around the torus in the mutually orthogonal planes corresponding to D and E. (I) Detailed structure of $\mathbf{n}(\mathbf{r})$ in the bulk (blue rods) and $\mathbf{n}_s(\mathbf{r})$ on the torus surface (black rods) induced by a $g = 1$ handlebody.

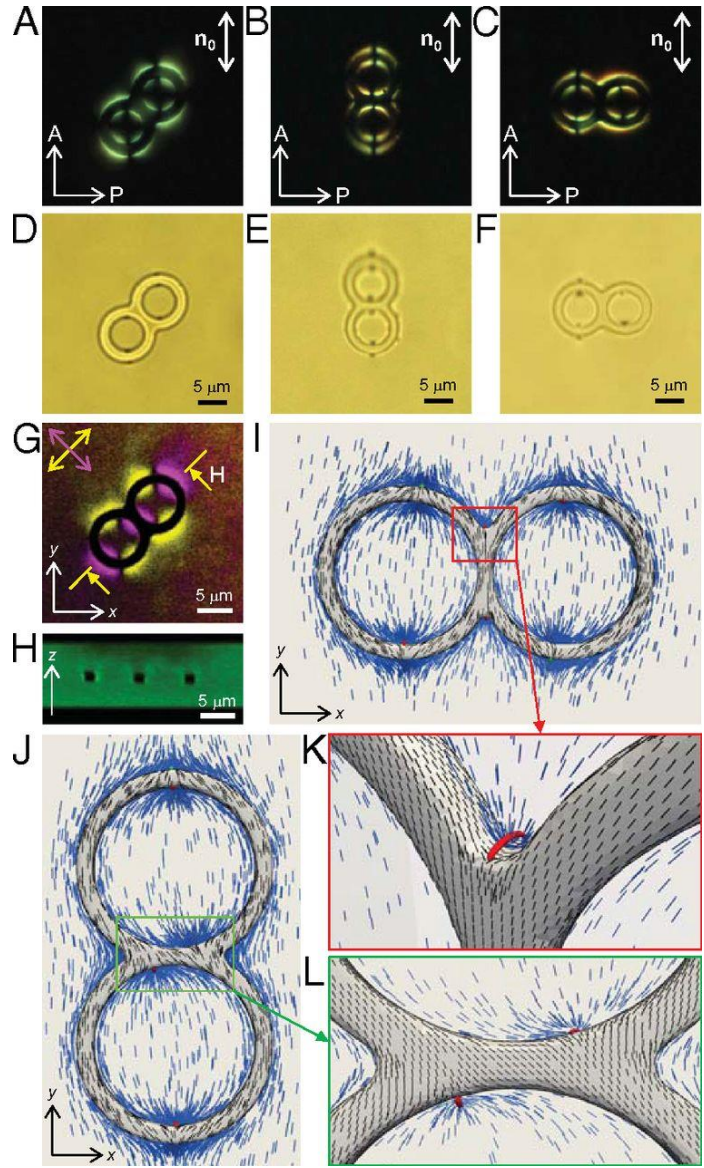

**Fig. 3.** Colloidal handlebody with $g = 2$ in a planar LC cell. POM (A–C) and bright-field (D–F) images of handlebodies having ring planes parallel to $\mathbf{n}_0$ but with the axis connecting centers of two rings at different orientations with respect to $\mathbf{n}_0$. (G) 3PEF-PM image obtained by superimposing 3PEF-PM textures taken for two orthogonal linear polarizations of laser excitation light marked by yellow and magenta double arrows. (H) 3PEF-PM vertical cross-section obtained along the H-H line marked in G. (I and J) Numerically calculated $\mathbf{n}(\mathbf{r})$ in the LC bulk (blue rods) and $\mathbf{n}_s(\mathbf{r})$ on the surface (black rods) of the $g = 2$ colloidal handlebody for two different orientations of its axis connecting centers of two rings with respect to $\mathbf{n}_0$. (K and L) Detailed views of $\mathbf{n}(\mathbf{r})$ and $\mathbf{n}_s(\mathbf{r})$ in the near-boojum regions marked in I and J, with the isosurfaces of constant scalar-order parameter $Q = 0.25$ shown in red.

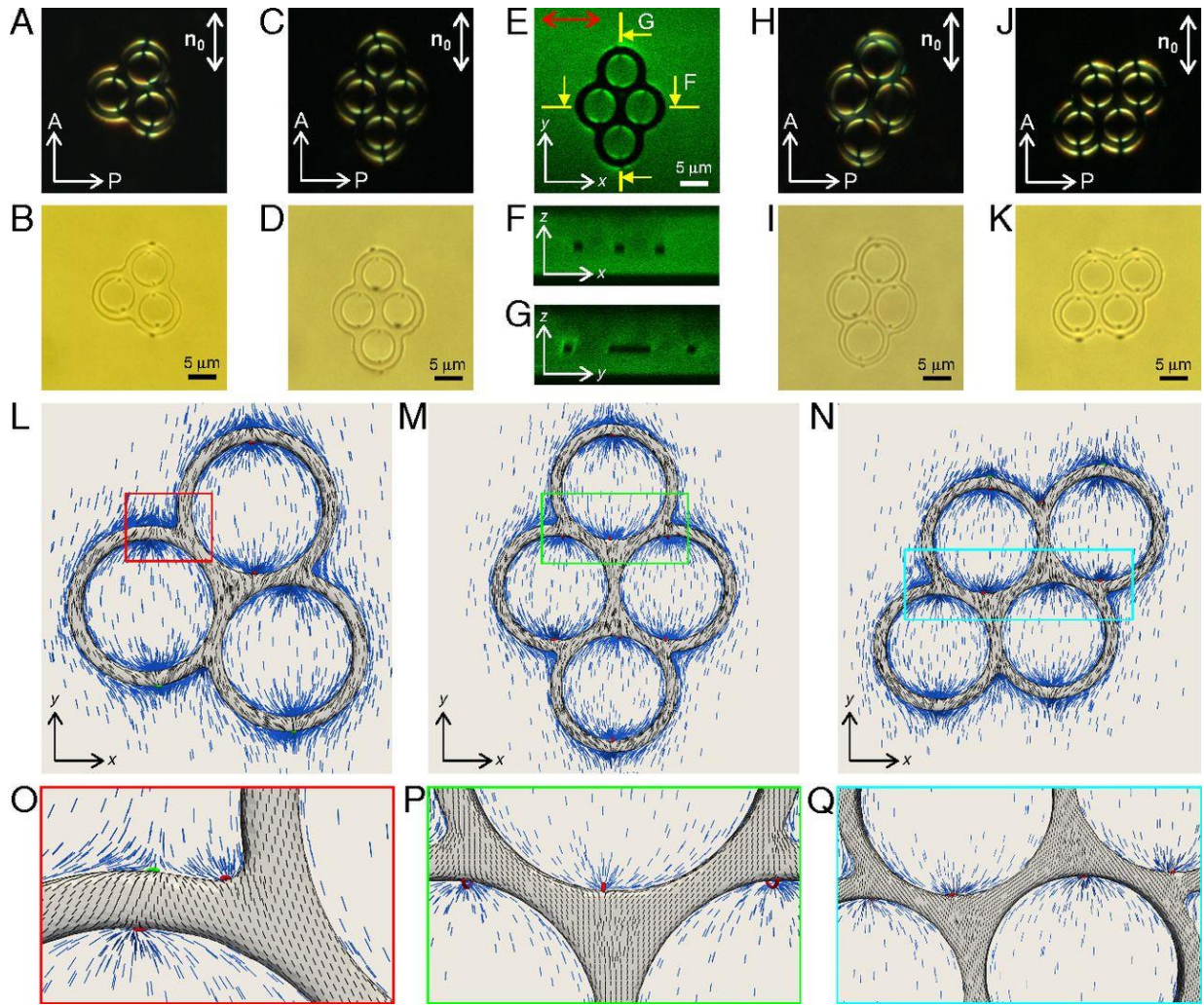

**Fig. 4.** Colloidal handlebodies with genus $g = 3$ and $g = 4$ in a planar LC cell. POM (A) and bright-field (B) images of a $g = 3$ particle. POM (C, H, and J), bright-field (D, I, and K), and 3PEF-PM (E–G) images of a $g = 4$ handlebody. (L–N) Numerically calculated $\mathbf{n}(\mathbf{r})$ in the LC bulk (blue rods) and $\mathbf{n}_s(\mathbf{r})$ on the surface (black rods) of the colloidal handlebodies at different orientations with respect to $\mathbf{n}_0$ corresponding to experimental images. (O–Q) Detailed views of $\mathbf{n}(\mathbf{r})$ and $\mathbf{n}_s(\mathbf{r})$ in the near-boojum regions marked in L–N. The isosurfaces of constant scalar-order parameter $Q = 0.25$ are shown in green and red.

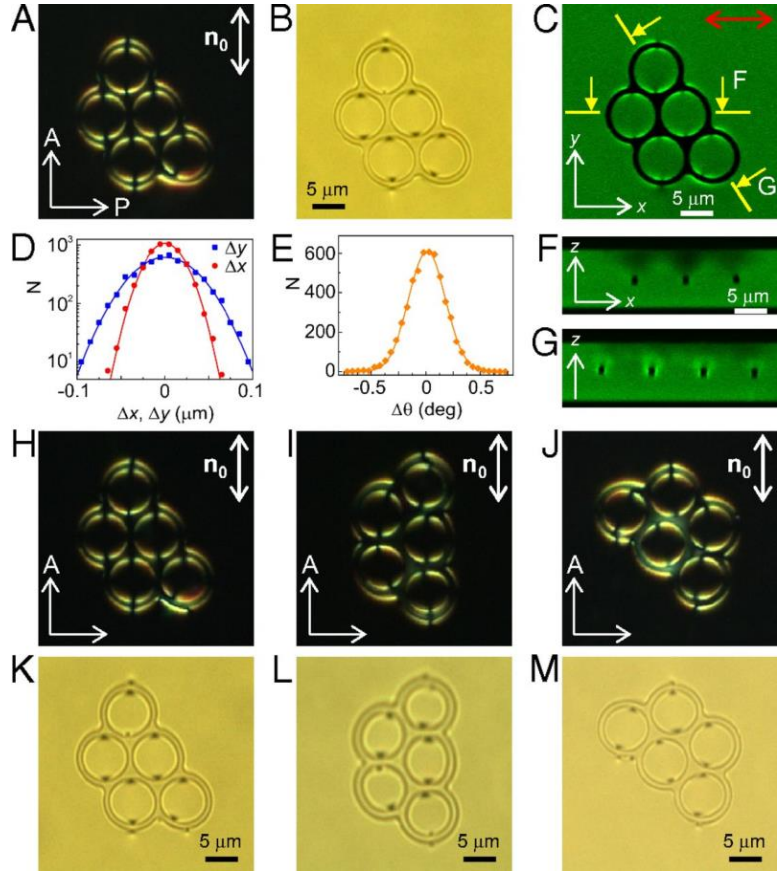

**Fig. 5.** Colloidal handlebody with $g = 5$ and ring planes parallel to $\mathbf{n}_0$. POM (A), bright-field (B), and 3PEF-PM (C) microscopy textures of a $g = 5$ handlebody in a planar LC cell. (D) Histogram for translational displacements of a $g = 5$ handlebody along the $x$ and $y$ directions (i.e., parallel and perpendicular to $\mathbf{n}_0$, respectively, as marked in C). (E) Histogram of the angular displacements of the same handlebody with respect to $\mathbf{n}_0$. Translational (D) and angular (E) displacements were characterized by probing Brownian motion using video tracking for at least 5 min, with an elapsed time of 0.067 s between frames. Experimental data (filled symbols) were fit with a Gaussian function (solid lines). (F and G) 3PEF-PM vertical cross-sections obtained along the F-F and G-G lines marked in C. POM (H–J) and bright-field (K–M) images of $g = 5$ handlebody colloids at different orientations with respect to $\mathbf{n}_0$ but all having their ring planes parallel to $\mathbf{n}_0$.

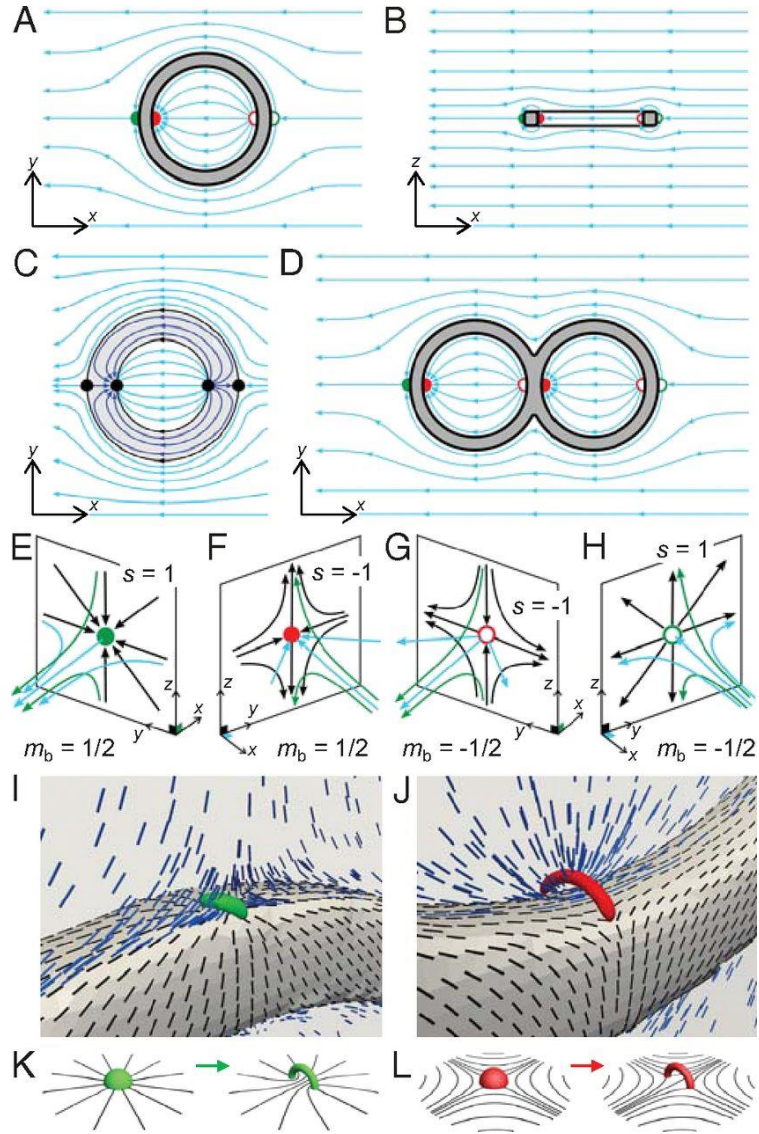

**Fig. 6.** Interplay of topology of boojums and colloidal handlebodies. Schematics of **n**(**r**) decorated with vector fields around a $g = 1$ handlebody in the ring plane (A) and in a plane orthogonal to the ring (B). (C) Schematic of the real vector field decorating **n**(**r**) around a colloidal torus (blue lines) and a virtual vector field inside the torus (dark blue lines), with real and virtual boojums being mirror images of each other and together comprising hyperbolic hedgehogs (black-filled circles) of different signs, demonstrating that the texture can be understood by treating boojums as halves of bulk hedgehog point defects. (D) Schematic diagram of **n**(**r**) decorated with vector fields around a $g = 2$ handlebody. (E–H) Detailed schematic diagrams of **n**(**r**) decorated with a vector field in the bulk (green and blue lines) and $\mathbf{n}_s(\mathbf{r})$ on the surface (black lines) near boojums having full topological charges $m_b = \pm 1/2$ and the corresponding disclination strengths in $\mathbf{n}_s(\mathbf{r})$ being $s = \pm 1$. (I and J) Numerically calculated **n**(**r**) and $\mathbf{n}_s(\mathbf{r})$ of the colloidal handlebodies in the vicinity of boojums. The isosurfaces of constant scalar-order parameter $Q = 0.25$ are shown in green and red. (K and L) Schematics show splitting of $s = \pm 1$ singularities in $\mathbf{n}_s(\mathbf{r})$ into pairs of $\pm 1/2$ and formation of handle-shaped disclination semiloops in the cores of boojums with radial (green) and hyperbolic (red) $\mathbf{n}_s(\mathbf{r})$.

**Table 1.** Disclination strengths and point defect charges due to *g*-tori in a nematic LC

| g | $\chi = \Sigma_i s_i = 2\Sigma_i N_{bi}$ | Number of defects with strength/charge | | | | Examples |
|---|---|---|---|---|---|---|
| | | $s = +1$ | $s = -1$ | $m_b = +1/2$* | $m_b = -1/2$* | |
| 1 | 0 | 2 | 2 | 2 | 2 | Fig. 1C |
| | | 0 | 0 | 0 | 0 | Fig. 2 |
| 2 | -2 | 2 | 4 | 3 | 3 | Fig. 3D |
| | | 4 | 6 | 5 | 5 | Fig. S3B |
| 3 | -4 | 2 | 6 | 4 | 4 | Fig. 4B |
| | | 4 | 8 | 6 | 6 | Fig. 4L |
| 4 | -6 | 2 | 8 | 5 | 5 | Fig. 4I |
| | | 4 | 10 | 7 | 7 | Fig. 4K |
| 5 | -8 | 2 | 10 | 6 | 6 | Fig. 5B |
| | | 3 | 11 | 7 | 7 | Fig. 5K |

*Characteristic of boojum $m_b$ is defined in the main text

## Supporting Information

It is convenient to define the dimensionless temperature $\tau = 24ac/b^2$. At $\tau < 1$ the uniaxial nematic is stable and the degree of orientational order is given by

$$Q_b = \frac{b}{8c}\left(1 + \sqrt{1 - \frac{8\tau}{9}}\right). \qquad [S1]$$

The nematic becomes unstable at $\tau > 9/8$. At $\tau = 1$ both the nematic and the isotropic phases coexist. We assume $a = a_0 (T - T^*)$, with $a_0$ a material-dependent constant and $T^*$ the supercooling temperature of the isotropic phase, and use typical values of the bulk parameters for 5CB are (1) $a_0 = 0.044 \times 10^6$ J/m$^3$, $b = 0.816 \times 10^6$ J/m$^3$, and $c = 0.45 \times 10^6$ J/m$^3$, $L_1 = 6 \times 10^{-12}$ J/m, $L_2 = 12 \times 10^{-12}$ J/m, $T^* = 307$ K. The spatial extension of inhomogeneous regions and the cores of topological defects is of the order of the bulk correlation length, which is given by $\xi = (8c \cdot (3L_1 + 2L_2)/b^2)^{1/2} \approx 15$ nm at the nematic-isotropic (NI) transition (2). The Landau-de Gennes elastic constants $L_1$ and $L_2$ may be related to the Frank-Oseen elastic constants (3, 4), $K_1 = K_3$ and $K_2$, through the uniaxial ansatz $Q_{ij} = (3/2) Q_b (n_i \cdot n_j - \delta_{ij}/3)$, yielding $K_1 = K_3 = 9 Q_b^2 (L_1 + L_2/2)/2$ and $K_2 = 9 Q_b^2 L_1 / 2$. In general $K_1$ and $K_3$ are different, but in most cases the difference is small and the LdG free energy is deemed adequate. We also introduce the dimensionless anchoring strengths $\omega = W Q_b^2 r / K_2$, where $r$ is some typical length scale related to the size of the colloidal particle and $W$ is an anchoring strength coefficient. For simplicity, we set $L_2 = 0$ in calculations.

The surface of a single toroidal particle is modeled implicitly in Cartesian coordinate as

$$\left(R - \sqrt{x^2 + y^2}\right)^{10} + z^{10} = r^{10}, \qquad [S2]$$

where $R$ is the distance from the center of the torus to the center of the tube, and $r$ is the tube radius. We assume $R/r = 10$, and use $R = 0.5$ μm $\approx 33\xi$, which gives $\omega \approx 19$. This surface is triangulated using Open Source GNU Triangulated Surface (GTS) library (6), the surfaces of

*g*-tori with *g* > 1 are constructed by using the set operation implemented in the GTS library. The triangulation of the nematic domain Ω is carried out using *Quality Tetrahedral Mesh Generator* (6), which supports the adaptive mesh refinement. Linear triangular and tetrahedral elements are used in 2D and 3D, respectively. Generalized Gaussian quadrature rules for multiple integrals (7) are used in order to evaluate integrals over elements. In particular, for tetrahedra a fully symmetric cubature rule with 11 points (8) is used, and integrations over triangles are done by using a fully symmetric quadrature rule with 7 points (9). The discretized Landau-de Gennes functional in then minimized using the *INRIA's M1QN3* (10) optimization routine, which implements a limited memory quasi-Newton technique of Nocedal (11).

**Supporting Figures**

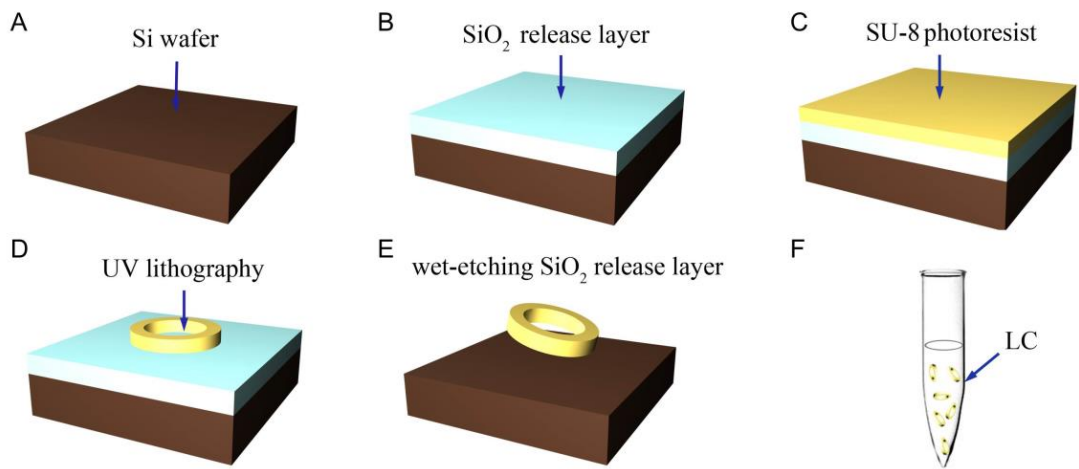

**Fig. S1.** Preparation of colloidal tori dispersed in LC. (A, B) We deposit a SiO$_2$ layer on a silicon wafer and then (C) spin-coat a layer of ultraviolet-sensitive photoresist SU-8. We then (D) define patterns of the tori by ultraviolet (UV) lithography and (E) wet-etch the SiO$_2$ layer to release the colloidal SU-8 tori. (F) The particles are then re-dispersed into a nematic LC after washing the particles in water and organic solvents as well as using solvent exchange and evaporation.

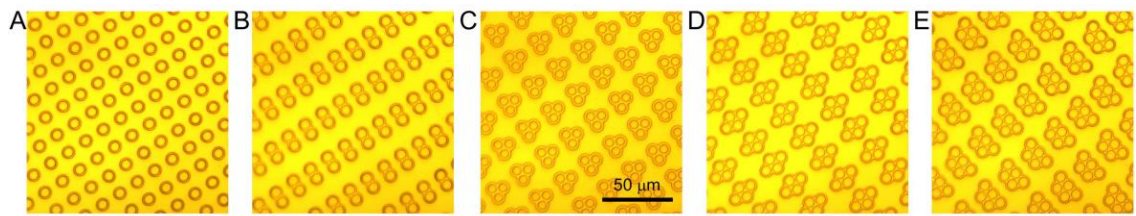

**Fig. S2.** Arrays of SU-8 colloidal tori on a silicon dioxide layer: (A) $\chi = 0$ and $g = 1$; (B) $\chi = -2$ and $g = 2$; (C) $\chi = -4$ and $g = 3$; (D) $\chi = -6$ and $g = 4$; (E) $\chi = -8$ and $g = 5$. Images were taken with bright field microscopy in reflection mode.

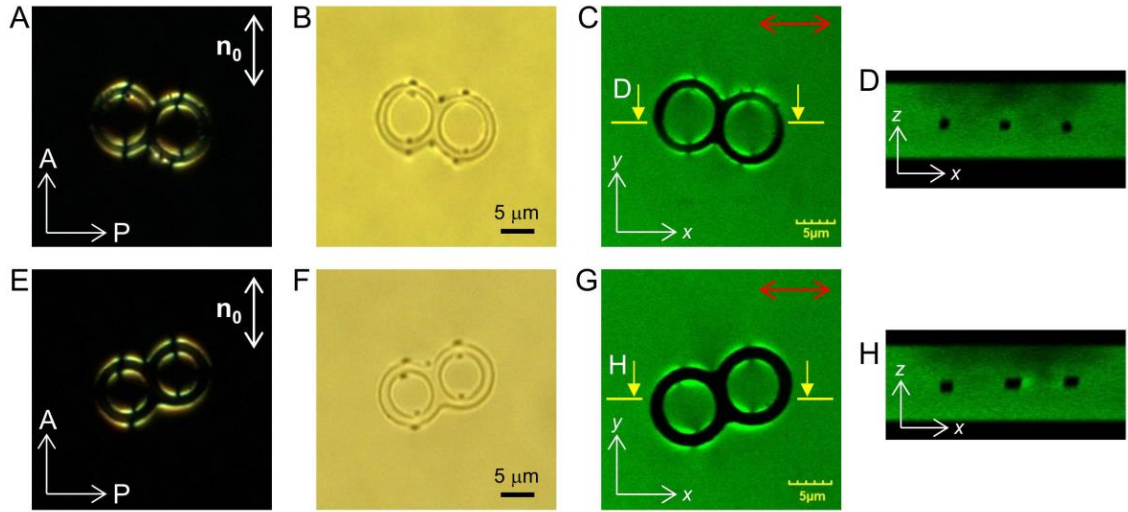

**Fig. S3.** Colloidal handlebody with $g = 2$ in a planar LC cell. (A, C) POM and (B, F) bright-field images of handlebodies having ring planes parallel to $\mathbf{n_0}$ but the axis connecting centers of two rings at different orientations with respect to $\mathbf{n_0}$. (C, G) In-plane 3PEF-PM images obtained at the linear polarizations of laser excitation light marked by red arrows. (D, H) 3PEF-PM vertical cross-sections obtained along the D-D and H-H lines marked in C and G.